Nick Paape*, Joost A. W. M. van Eekelen and Michel A. Reniers

# Automated design space exploration for poultry processing systems using discrete-event simulation

**Abstract:** The poultry processing industry struggles to keep up with new developments in meat consumption and livestock breeding. Designing poultry processing systems is becoming increasingly more complex, and an increasing number of iterations of (re)design are required to optimize the product flow in these systems. To address this issue, this study presents a discrete-event simulation-based method for design space exploration of production systems. This method is mostly automated, greatly reducing the time and effort required in the design process. The steps that are automated are iterating on the design, model construction, performing simulation experiments, and interpreting the simulation results. An industrial case study in which a poultry processing system is redesigned is used to validate the effectiveness of the proposed method. A detailed description of this case study is given to showcase the different ways in which this design space exploration method can be used.

## 1 Introduction

### 1.1 Problem statement

The meat processing industry is undergoing rapid changes. One example is the growing mismatch between providers and consumers in the poultry industry. On the one hand, broilers (chickens bred for meat consumption) are being bred increasingly heavier [1]. On the other hand, consumer purchasing habits are shifting [2]; lighter chicken fillets are preferred, as they are perceived as more sustainable. Dealing with these developments requires new methods for designing meat processing systems.

The above mismatch introduces extra complexity in the design process of a poultry fillet processing system. The fillet processing system must match the incoming flow of predominantly heavy fillets to production orders, which require mostly light fillets. This problem can be partially alleviated by optimizing the product flow in the layout of the fillet processing system. Another more recent solution is the introduction of new 'trimming' machines, which can trim down heavy fillets into lighter fillets, and a by-product that can be used in minced (ground) poultry products. However, the introduction of these machines also leads to new design questions, such as how many of these machines are to be used in a layout and where in the layout should these machines be placed. More design choices lead to a larger number of feasible designs, and new methods are needed to help a designer explore the ever-growing solution space. The solution space of feasible designs can be referred to as a system's *design space* [3]. The design process for poultry processing systems, as well as production systems in general, is frequently a difficult task requiring numerous iterations of design and redesign. What further complicates the design process, is that these systems generally have multiple performance objectives. Identifying which designs are optimal requires multi-objective optimization [4].

In this work, a 'design' is regarded as the functional architecture of the system, which means that a design describes the machines of the system and the connections between them. Integrating simulation in the design process allows for design alternatives to be compared, such as shown for an extruded food production line in Owens and Levary [5]. However, using simulation in the design process also introduces additional steps. It requires the system designer to specify the different designs, construct the models of these designs, perform simulation experiments on these models, and interpret the simulation results, only to repeat these steps in the next design iteration.

Ideally, the design of a production system is optimized automatically with respect to the selected performance

*Corresponding author: Nick Paape, Eindhoven University of Technology, PO Box 513, Eindhoven, 5600 MB, The Netherlands,
E-mail: n.paape@tue.nl. https://orcid.org/0000-0001-8835-525X
Joost A. W. M. van Eekelen and Michel A. Reniers, Eindhoven University of Technology, PO Box 513, Eindhoven, 5600 MB, The Netherlands

objectives. One area of research that can help in this regard is that of *design space exploration*, which discusses how design alternatives can be automatically analyzed and compared [3]. In this paper, a method is proposed for automated design space exploration, in which discrete-event simulation is used for multi-objective optimization of the system design. The proposed method is demonstrated on a poultry processing system, but can be applied to other food processing systems, and to production systems in general.

## 1.2 State of the art

This section provides overviews of applications of simulation in the design of food processing systems and of methods for (discrete-event) simulation-based design space exploration of production systems in general.

### 1.2.1 Application of simulation in the design of food processing systems

Simulation is used to design an industrial plant for the production of hazelnut-based products in Bruzzone and Longo [6]. A simulation model is used to analyze the behavior of the designed system in various operating conditions. Penazzi et al. [7] shows how simulation can be used in the design and control of food job-shop processing systems. This is demonstrated through a case study in the catering industry, in which simulation is used to analyze a collection of key performance indicators.

In Owens and Levary [5] simulation is used to analyze design alternatives for an extruded food production line. The authors demonstrate how simulation may be a useful decision-support tool for deciding between several design options in the food industry. Parthanadee and Buddhakulsomsiri [8] shows how value stream mapping and simulation can be used to analyze design choices for small and medium enterprises in roasted ground coffee production.

Discrete-event simulation is used in Plà-Aragonés et al. [9] to examine alternative processing strategies and to improve production planning for a pig meat packaging facility. Comparing discrete-event simulation to deterministic or stationary methods, Plà-Aragonés et al. find that discrete-event simulation better captures plant behavior. Discrete-event simulation is used in Rijpkema et al. [10] to redesign meat production processes to make more effective use of product quality information. Simulation was used to evaluate how sorting based on product quality information affects the performance and processing efficiency of the system. As shown by Plà-Aragonés et al. and Rijpkema et al., discrete-event simulation can be an effective tool in the design of meat processing systems.

In Fujii et al. [11] simulation is used to optimize the design of the facility layout of a central kitchen in the food service industry. The proposed method for layout planning is validated through a real-scale case study. In Masoud et al. [12] a simulation-based framework is developed to optimize the facility layout and resource allocation in vegetable grafting facilities in order to maximize production capacity and efficiency. Discrete-event simulation is used in the design process of a greenhouse for industrial head lettuce production in Gao et al. [13]. A framework is proposed for integrating systematic layout planning and simulation to design greenhouses that maximize efficiency and yield.

When designing a food processing system, the first goal is always to meet the specified requirements; is the system able to produce or process the products as intended? A secondary goal is often to optimize the design in regard to certain performance measurements such as throughput and cost. Simulation allows the capabilities of potential designs to be evaluated early in the design process, across a range of production scenarios. Bruzzone and Longo [6], Penazzi et al. [7], and Plà-Aragonés et al. [9] show how simulation can be used to evaluate the performance of a food processing system across a range of production scenarios. Owens and Levary [5], Fujii et al. [11], and Gao et al. [13] demonstrate how simulation can be used to compare alternative designs in the design process of food processing systems. Rijpkema et al. [10], Parthanadee and Buddhakulsomsiri [8], and Masoud et al. [12] combine both, and show the importance of evaluating multiple production scenarios when using discrete-event simulation to compare alternative designs. Finally, Fujii et al. [11] and Masoud et al. [12] show a form of automated design space exploration to solve the facility layout planning problem. However, in both methods, the design is regarded as the spatial layout of the system, with the goal being to minimize the distance that products or workers need to travel. To the best of our knowledge, there is no previous work on automated design space exploration for the functional architecture of food processing systems.

### 1.2.2 Using discrete-event simulation for design space exploration

A digital factory is created using a discrete-event simulation tool in Centobelli et al. [14]. This digital factory is utilized to optimize the facility layout with respect to the flow of material. Simulation is used to compare the current and proposed layout for a range of production scenarios. An algorithm for generating a discrete-event simulation model of an assembly system layout is proposed in Kranz et al. [15]. This algorithm allows for simulation models to be generated without the need for considerable expertise in simulation. The model is generated based on predefined process logic and layout data provided in an Excel spreadsheet.

A method for automatic simulation model generation of a robotic cell layout is provided in Laemmle and Gust [16]. The method uses layout data specified in AutomationML as input. The facility layout of a specialized furniture manufacturing plant is optimized using discrete-event simulation in Rodič and Kanduč [17]. The facility layout is optimized to minimize the total travel distance for products. Simulation models of the different facility layout designs are automatically generated.

The downside of using discrete-event simulation in the design process of a production system is that developing the models can be complex and time-consuming. Kranz et al. [15], Laemmle and Gust [16], and Rodič and Kanduč [17] overcome this downside by using automated model construction. Rodič and Kanduč [17] automates both model construction and design space exploration to solve the facility layout planning problem. However, none of these works uses automated model construction to automatically explore the design space of the functional architecture of a production system.

In all of the papers previously mentioned in the state of the art, simulation is either used to directly compare design alternatives, or is used for single-objective optimization. None of these methods allow for multi-objective optimization, even though there is often more than one performance measure that needs to be considered in the design process of a production system.

## 1.3 Contribution

This paper presents a method for design space exploration of production systems. The novelty of our contribution is in how the individual steps listed below are combined to automate the design space exploration process, and in demonstrating how this proposed method can be applied to a case study of industry-level complexity in the poultry processing industry. The proposed method is explained and validated through an industrial case study of a poultry fillet processing system. Table 1 shows a comparison of the proposed method to the state of the art. In the proposed method:
– The design space of the proposed system is explored iteratively & automatically.
– The (discrete-event simulation) models for the designs are automatically constructed using a model library.
– These models are used to carry out simulation experiments to predict the system's performance in a predefined set of production scenarios (a product scenario describes the conditions under which the production system operates).
– The system designs are then evaluated across these production scenarios, for the chosen performance objective(s). Both single- and multi-objective optimization of the design are supported.

This work is a continuation of the work presented in Paape et al. [18]. This work presents an in-depth explanation of the proposed design space exploration method and an extension of the case study. A detailed description is given of:
– Specification of the design space using a 'design space matrix'.
– The development of the model library in Anylogic [19].
– How the production scenarios that are simulated were chosen.
– The trade-off between accuracy and computation time, and the tuning of simulation parameters such as simulation duration, warmup period, and the number of independent simulation runs.
– Examples of how the design space can be explored using single- or multi-objective simulation.

The structure of this paper is as follows. First, in Section 2, the motivating case study is described. Then, in Section 3, the proposed method for design space exploration is presented. The steps of the proposed method are explained using the case study. Next, Section 4 features a critical discussion on the advantages and limitations of the proposed method. Finally, Section 5 concluding remarks are made, along with ideas for future work.

## 2 Case study description

A case study was conducted on an existing poultry plant. This case study is used to showcase the proposed design space exploration method of Section 3. The plant was redesigned to better match the predominantly heavy chicken fillets which are harvested from the broilers, to production orders requiring mostly light fillets. Figure 1 shows the parts of a poultry processing plant that are relevant to the processing of fillets. First, the broilers arriving at the plant are distributed over a number of cut-up lines. These cut-up lines harvest (among others) the fillets, which are then sent downstream to the fillet processing (sub)system. The purpose of the fillet processing system is to distribute the fillets of different weights to the most suitable downstream destinations. How these fillets are distributed is decided by a production controller, and depends on the layout of this subsystem. The downstream destinations are where the ᵃIt is unclear which type of simulation is used.
ᵇThe design space concerns the spatial layout of the production system.
ᶜThe design space concerns the functional architecture of the production system.

Table 1: The proposed method compared to the state of the art.

| | Owens and Levary [5] | Bruzzone and Longo [6] | Penazzi et al. [7] | Parthanadee and Buddhakulsomsiri [8] | Plà-Aragonés et al. [9] | Rijpkema et al. [10] | Fujii et al. [11] | Masoud et al. [12] | Gao et al. [13] | Centobelli et al. [14] | Kranz et al. [15] | Laemmle and Gust [16] | Rodič and Kanduč [17] | This paper |
|---|---|---|---|---|---|---|---|---|---|---|---|---|---|---|
| Discrete-event simulation | ✓ | ✓ | ✓ | ✓ | ✓ | ✓ | ?ᵃ | ✓ | ✓ | ✓ | ✓ | ✓ | ✓ | ✓ |
| Comparison of design alternatives | ✓ | | ✓ | ✓ | | ✓ | ✓ | ✓ | ✓ | ✓ | ✓ | ✓ | ✓ | ✓ |
| Evaluation across multiple scenarios | | ✓ | ✓ | ✓ | | ✓ | | ✓ | | ✓ | | | | ✓ |
| Automated design space exploration | | | | | | | ✓ᵇ | ✓ᵇ | | | | | ✓ᵇ | ✓ᶜ |
| Automated model construction | | | | | | | | | | | ✓ | ✓ | | ✓ |
| Application in food processing | ✓ | ✓ | ✓ | ✓ | ✓ | ✓ | ✓ | ✓ | ✓ | | | | ✓ | ✓ |
| Single or multi-objective optimization | | | | S | | | S | S | | S | | | S | M |

different end products are produced, such as batches of fillets in trays and chicken burgers. Finally, these end products are packaged and sent to customers.

In the case study, the fillet processing system was redesigned. The performance of the system is measured by how well it is able to meet the throughput targets for the given production orders of end products such as batches and burgers. To properly predict the system's performance, requires that a wide range of production scenarios is analyzed so that the variation in production orders throughout the year is aptly captured. Some modifications have been made to the case study to ensure confidentiality. However, the complexity of the modified case study is similar to that of the original when it comes to the size of the design space. The first difference is in the plant layer, where different (but similar) numbers of lanes, destinations, and modules are used. However, the types of physical modules are the same as in the original case. The second difference is in the production controller; in the actual system, a more elaborate production strategy is utilized for more optimal performance.

First, the characteristics of the inflow of fillets arriving from the cut-up lines are described in Section 2.1. Next in Section 2.2, 'recipes' are explained, which specify how production orders must be fulfilled. Finally, a description of the system is given in Section 2.3.

## 2.1 Fillet inflow

Each day a poultry processing plant processes multiple different 'flocks' from various farmers. Each flock has its own characteristic weight distribution. The weight of the flock can have a large influence on which orders can be fulfilled, and thus on the performance of the system. Poultry fillet weights are generally normally distributed [20]. For this case study, it is assumed that the mean fillet weight of a flock is between 200 and 300 g, with a standard deviation equal to 10 % of the mean weight.

In a poultry processing plant, the broilers in a flock are distributed over multiple cut-up lines according to their weight to maximize production yield. For this case study, it is assumed that each of the five lanes of the poultry fillet processing system receives a quintile of the weight distribution. For example, the first lane in Figure 3 processes the lightest 20 % of fillets, and the fifth lane processes the heaviest 20 %. An example of the weight distributions per lane for a flock with a mean weight of 250 g is shown in Figure 2.

## 2.2 Recipes

A poultry processing system has a set of production orders it aims to fulfill, for end products such as trays of fillets, chicken schnitzels, and chicken burgers. There are requirements for what fillet weights can be used for each end order. For example, to make chicken schnitzels only fillets of between 200 and 350 g can be used. The requirements of which products can be used, and how they should be processed, are described in a *recipe*. A recipe describes:
– The product destination of products for this recipe.
– The priority of the recipe.
– The target throughput in fillets/minute.
– A lower and upper limit for fillet weight (post-trim).
– A weight limit for trimming (most customers want to limit how much of a fillet is trimmed).

The production targets of a poultry processing plant are season-dependent. A good design for a fillet processing system performs well throughout the entire year. In this case study, typical production orders for summer and for winter are analyzed. The sets of recipes for the two seasons can be seen in Table 2. In both instances, there is a default recipe for fillets that do not fit any of the other recipes.

## 2.3 System description

A fillet processing system can be decomposed into a (physical) plant layer and a (cyber) production control layer. In

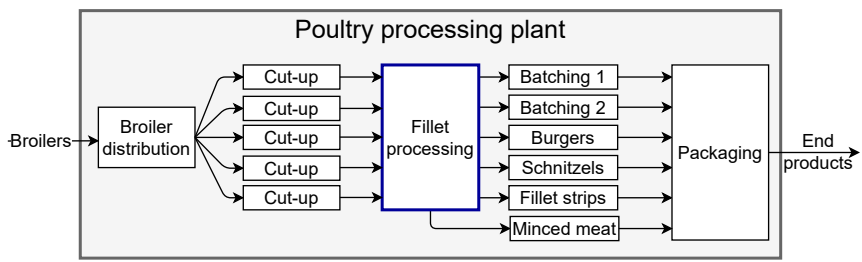

**Figure 1:** The parts of a poultry processing plant that are relevant to the processing of fillets. The case study in this paper focuses on the (highlighted) fillet processing system. This system receives fillets from the five cut-up lines and distributes them to various subsystems that use them to produce end products. If the fillet processing system has the capacity for trimming down fillets to a lighter weight, then the by-product is sent to the minced meat subsystem.

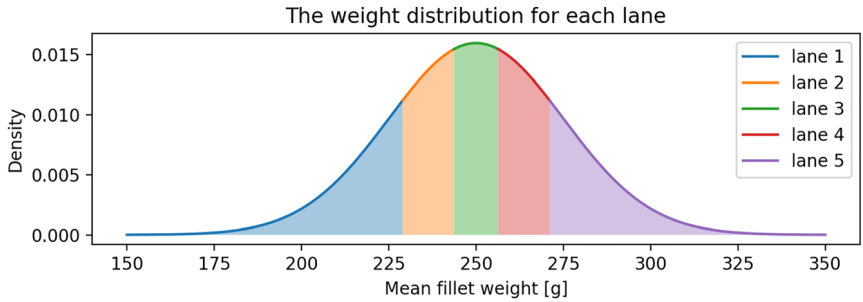

**Figure 2:** The weight distribution of each lane for a mean fillet weight of 250 g.

**Table 2:** The typical recipes for the summer and winter production orders.

| | | | Summer | | | |
|---|---|---|---|---|---|---|
| Recipe | Destination | Priority | Target throughput (fillets/min) | Min. fillet weight (g) | Max. fillet weight (g) | Max. trim weight (g) |
| 1 | Batching 1 | 1 | **130** | 100 | 250 | 100 |
| 2 | Batching 2 | 2 | **130** | 150 | 300 | 100 |
| 3 | Burger | 3 | **30** | 200 | 350 | 0 |
| 4 | Schnitzel | 4 | **30** | 250 | 400 | 0 |
| Default | Fillet strips | – | – | 0 | 1000 | 0 |
| | | | Winter | | | |
| Recipe | Destination | Priority | Target throughput [fillets/min] | Min. fillet weight [g] | Max. fillet weight [g] | Max. trim weight [g] |
| 1 | Batching 1 | 3 | **100** | 100 | 250 | 100 |
| 2 | Batching 2 | 4 | **100** | 150 | 300 | 100 |
| 3 | Burger | 1 | **60** | 200 | 350 | 0 |
| 4 | Schnitzel | 2 | **60** | 250 | 400 | 0 |
| Default | Fillet strips | – | – | 0 | 1000 | 0 |

In this example, only the throughput and priority are changed (highlighted in bold). However, other recipes with different values in any of the columns are also possible.

this subsection, the two layers will be explained. The current design of the fillet processing system is shown in Figure 3.

### 2.3.1 Plant layer

The plant or physical layer of the system consists of a set of production 'lanes'. Each lane represents a conveyor system through which the fillets are transported, with production steps occurring in-line. The plant layer can be decomposed into modules that define the production steps carried out on the fillets. The origin and destination modules represent the inflow from and outflow to other subsystems. For this case study, it is assumed that the order of processes is always the same (in the order in which they are described below). The 'modules' of the plant layer operate as follows:

#### 2.3.1.1 Origin
The origin module is where the fillets flow into the fillet processing system. Each lane has a different segment of the total weight distribution, as explained in Section 2.1.

#### 2.3.1.2 Weighing
The first production step is weighing the fillets. The measured product weight is sent to the production controller. The production controller uses data on the fillet weight distribution at each lane to calculate a production strategy (more on this in the next subsection).

#### 2.3.1.3 Assignment
After the fillets are weighed, a request for the product to be assigned is sent to the production controller. The controller then decides on the product routing and trim instruction based on the calculated production strategy. A product routing determines to which destination the fillet should be distributed. A trim instruction describes if and how much of the fillet should be trimmed.

#### 2.3.1.4 Trimming (optional)
Some lanes have a trimming station, which can be used to trim a small piece of a fillet. This is done to make 'heavy' fillets suitable for recipes that need 'light' fillets. How a fillet is trimmed depends on the assigned trim instruction.

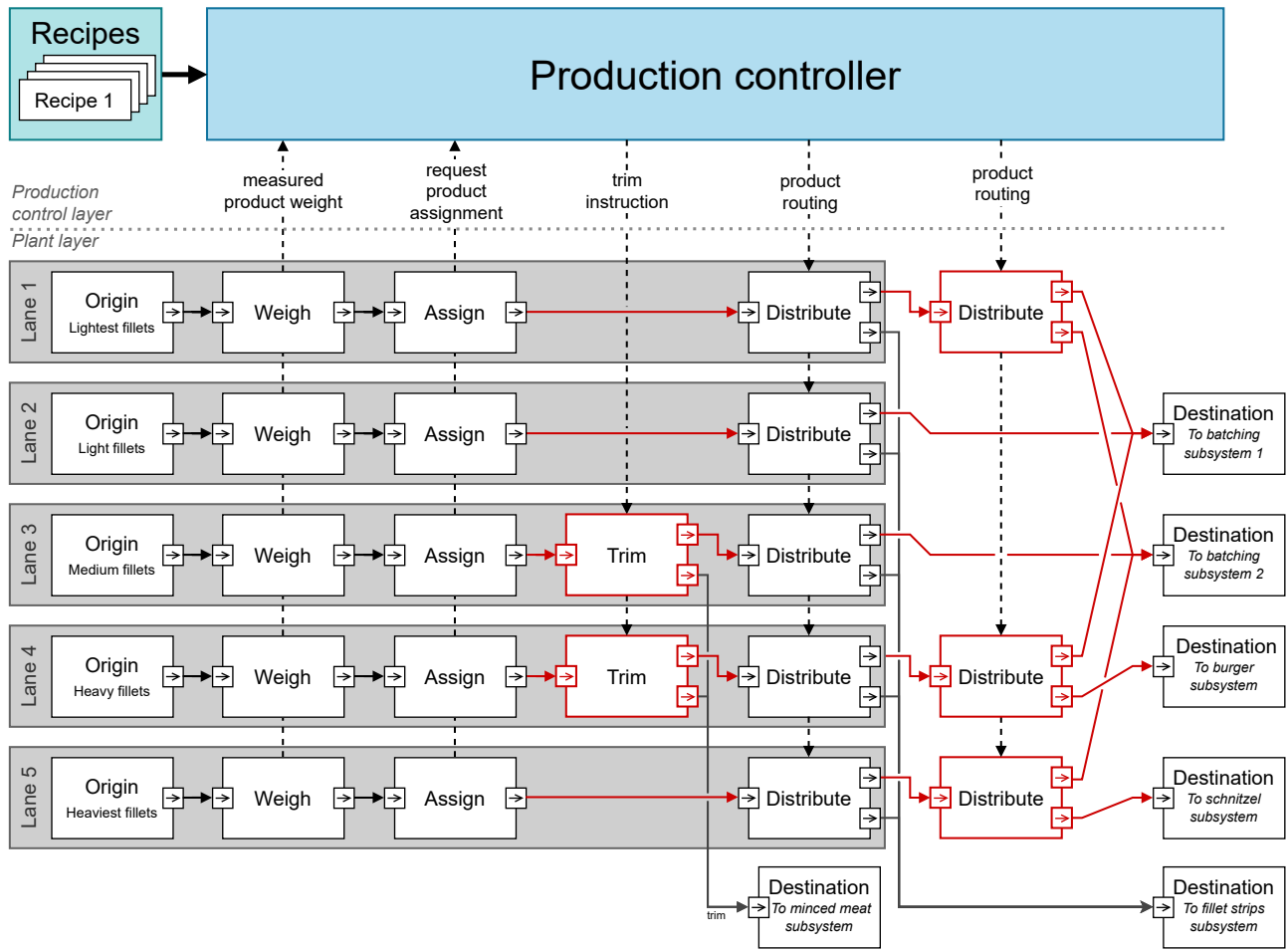

**Figure 3:** The architecture of the current fillet processing system. The system can be divided into a plant layer and a production control layer. In red are the connections and modules which can be modified in the redesign.

The trimmed by-product is sent to the 'minced meat subsystem' destination where it is used to make minced poultry products.

### 2.3.1.5 Distribution

The last oduction step in the fillet processing system is distribution to the different destinations. Due to technological constraints, product flows can only diverge, not merge. The exception is when fillets are sent to the fillet strips destinations. Distributors are used to diverge product flow in one of two directions. Each lane requires at least one distributor with a connection to the 'fillet strip subsystem', this is the default destination for fillets that cannot be used for other end products. Besides these five fixed distributors, there are three distributors to be placed freely.

### 2.3.1.6 Destination

After distribution products arrive at their destination. In a fillet processing system, the destinations are the starting points of subsequent subsystems, of which there are multiple types (e.g., for producing batches, burgers, schnitzels, etc.). Each destination has its own recipe, with different requirements for the weight of fillets. As mentioned earlier, the 'fillet strip subsystem' destination is a special destination which is the default destination. The 'minced meat subsystem' destination is another special destination for the produced trim.

In this case study, there are two design questions that are to be analyzed. First, how many trimming modules are needed, and where should they be placed? These trimming modules are costly, and the client would like to know what the added value of each additional trimming module is, and in which lanes these trimming modules should be placed. Second, which lanes should be connected to which destinations, and how should the three extra distributors be used? The modules and connections of the current design which can be modified are highlighted in red in Figure 3.

### 2.3.2 Production control layer

The production controller is responsible for ensuring that the target throughput of each of the recipes is met. It does so by calculating a production strategy. A production strategy describes which fillet weights are allocated to which recipes and which fillets must be trimmed. When calculating the production strategy, the production controller takes into account which lanes are connected to which destinations, and which lanes have the capability for trimming. The interactions between the production control layer and the plant layer are shown in Figure 3. The production controller functions as follows:

(1) The production controller collects data on the weights of fillets in each lane, as measured at the weigher in each lane.
(2) The production controller employs a sliding window technique, consisting of the most recent $N$ fillet weight measurements from each lane, to generate a 'measured fillet weights' histogram. This histogram describes the measured arrival rate of the different fillet weights in each lane. An example of such a histogram is shown in the top figure of Figure 4.
(3) The production controller calculates the production strategy every $t$ seconds using Algorithm 1. This algorithm functions as follows:
  (a) The recipe with the highest priority is selected.
  (b) The controller determines which lanes have a route to the destination of the recipe. The 'measured fillet weights' histograms of these lanes are used to calculate the expected throughput of each fillet weight interval.
  (c) Starting from the lower weight limit of the recipe, the weight range allocated to this recipe is increased, until either the upper weight limit is reached, or until the expected throughput equals the target throughput of the recipe.
  (d) If the target throughput is not yet reached, then the controller determines which of the selected lanes have trimming modules.
  (e) Starting from the upper weight limit of the recipe, the weight range allocated for trimming is increased, until either the upper weight limit for trimming is reached, or until the expected throughput equals the target throughput of the recipe.
  (f) The calculated weight ranges are assigned to the recipe. These weight intervals can no longer be used in other recipes. The process is then repeated for the next recipe until all recipes are processed.
  (g) The middle figure of Figure 4 shows an example of a production strategy and the expected throughput allocated to each recipe. Fillets between 100 and 300 g are allocated to batching, with fillets between 250 and 300 g being trimmed. Fillets above 300 g are used for schnitzels.
(4) Upon the arrival of a product at an assignment module, the production controller assigns a product to a recipe, and determines the required product routing and trim instructions. This is based on the production strategy for that lane and the measured weight of that fillet.
(5) Whenever a product arrives at a distribution or trimming module, the corresponding product routing or trim instruction is communicated to that module for execution. An example of the resulting production output is shown in the bottom figure of Figure 4.

---

**Algorithm 1:** production strategy calculation

```
 1  repeat
 2      Select (next) recipe with highest priority.
 3      Select lanes with a route to destination of recipe.
 4      Starting from lower weight limit of recipe:
 5      repeat
 6          Increase the weight range.
 7          Calculate expected throughput for selected lanes.
 8      until (target throughput is reached) OR (recipe's upper weight limit is reached)
 9      if target throughput is not yet reached then
10          Identify selected lanes with trimming.
11          Starting from upper weight limit of recipe:
12          repeat
13              Increase the weight range for trimming.
14              Calculate expected throughput for selected lanes.
15          until (target throughput is reached) OR (recipe's trimming weight limit is reached)
16      Assign (available) weight intervals to selected lanes.
17      Make weight intervals unavailable for next recipes.
18  until all recipes are processed
```

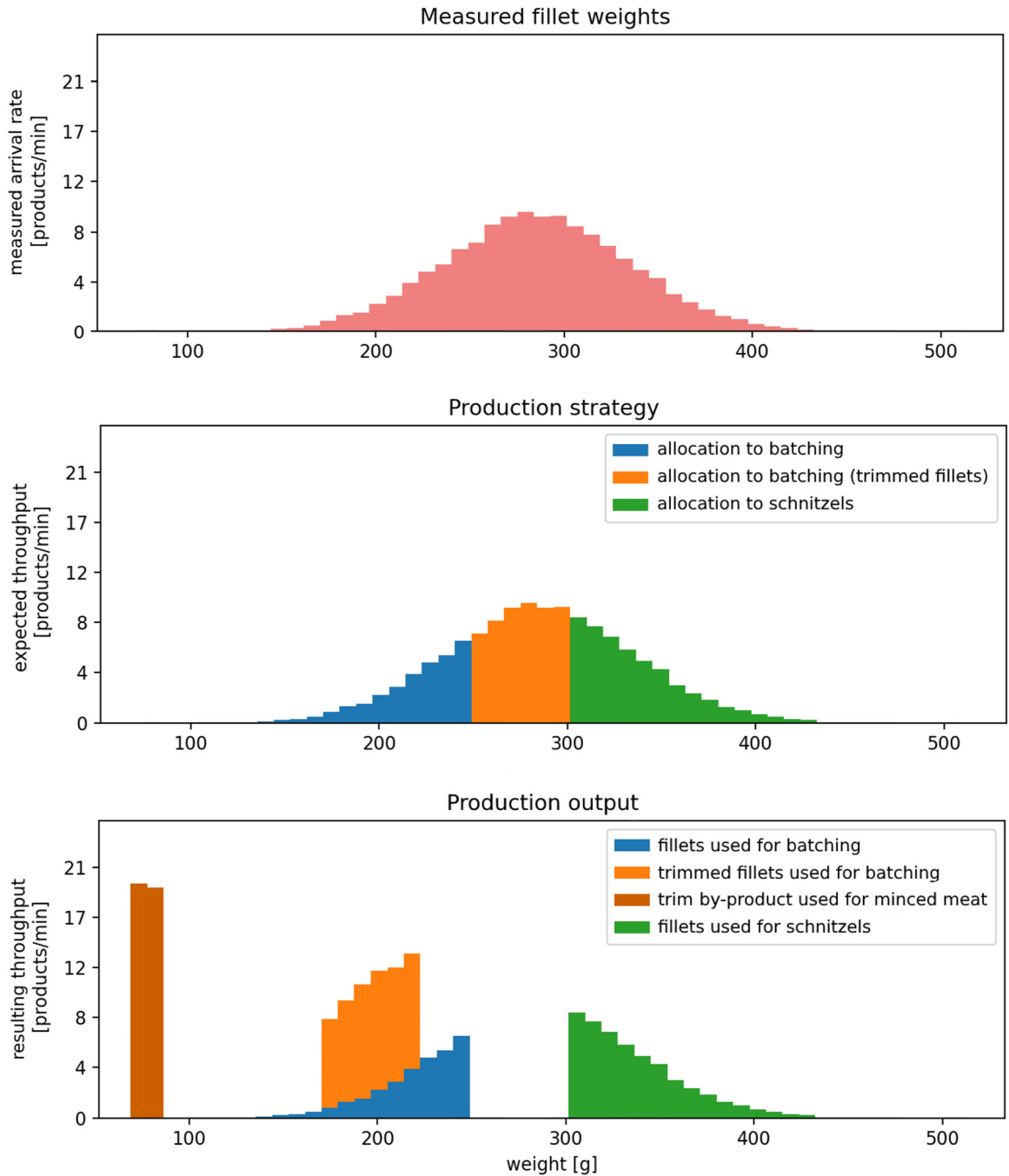

**Figure 4:** An example of the production strategy for a system with one lane, with recipes for batching and schnitzels. The top histogram shows the arrival rate for different fillet weights, as measured at the weigher. The middle histogram shows the production strategy as calculated using Algorithm 1. The bottom histogram shows the resulting production output. Fillets between 250 and 300 g are trimmed down to around 200 g so that they can be used for the batching recipe. The trim by-product is used in the production of minced meat.

# 3 Proposed method for design space exploration

This section explains the proposed method for design space exploration, which is shown in Figure 5. This method can be used to automatically explore the design space of a production system using discrete-event simulation. The proposed method uses the following four steps: *design space exploration*, *model construction*, *simulation*, and *evaluation*. After iterating through all designs, the method outputs a set of recommended designs. There can be multiple recommendations, from which the system designer can select the most suitable option. Details on these steps, as well as the input and output for each step, will be explained further below, using the case study of the fillet processing system as an example. The use of the method is not limited to fillet processing systems but should extend to production systems in general.

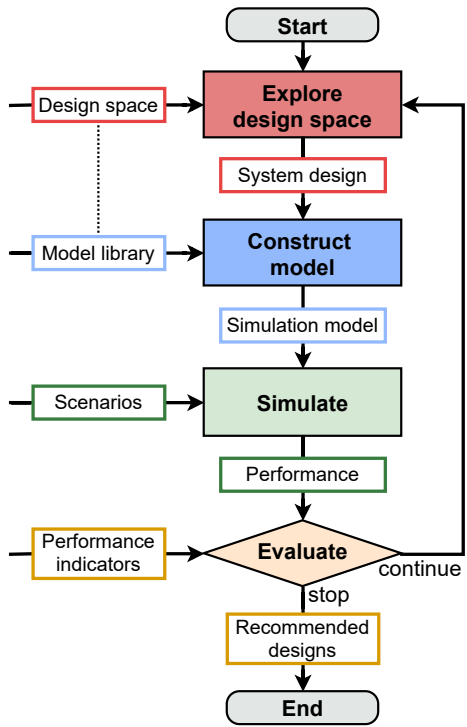

**Figure 5:** The method for design space exploration.

## 3.1 Design space exploration

The first step in the design space exploration of a production system is defining the design space itself. A design is described by a set of modules, with connections between their input and output ports. The design space can be considered the set of allowed modules, and the possible ways in which these modules can be connected. In this methodology, the design space is defined through the use of a Design Space Matrix (DSM), which describes all permissible connections between the output and input ports of the selected modules. The columns and rows of the DSM represent the allowed connections of each input and output port respectively. An element $(i, j)$ of the DSM is a Boolean value that indicates whether the output port $i$ of a module can be connected to the input port $j$ of another module. After specifying in the DSM which connections are allowed, the minimum and maximum number of connections that each port can have must also be specified. If a module can have both 0 input and output connections, then this means the module can be excluded from a design.

The design space exploration process begins by iterating through the various configurations of the DSM. In each iteration, a new design is generated. For every design, each port is connected between the minimum and maximum number of allowed connections for that port. It is possible that some modules may not be connected in the generated configuration of the system design. This allows for the evaluation of the performance of a system with and without specific modules.

### 3.1.1 Case study

The design space for the case study is presented in Table 3. In the table, an example of a possible system design is highlighted. In the fillet processing system, the parameters of the design space are: which lanes have a trimming module, and how are the lanes connected to the destinations (using the three distributors which can be placed freely). Most connections in the system are fixed. For example, 'origin1.out' must be connected to 'weigh1.in'; it is the only possible connection for both input and output ports, and both ports must have exactly one connection. In this specific case study, it is required that either all or none of the ports of a module are connected. Configurations of the system design that do not satisfy this condition are discarded. The case study design space is represented by a Design Space Matrix of $42 \times 30$, from which 11,520 possible configurations of the system design can be derived (32 permutations for which lanes have trimming modules, multiplied by the 360 permutations for how lanes can be connected to destinations).

## 3.2 Model construction

The next step in this method is the 'model construction' step, in which a simulation model of the system design is constructed using a model library in a modeling and simulation environment. This requires that all system modules have a corresponding model component (Figure 5 shows this relationship between the design space and the model library). It also requires that these model components are modeled modularly, and that the model components can be connected dynamically. Finally, model construction requires that the production control layer of the system can be automatically adjusted to different system designs.

When constructing the model library, it is important to define if and what performance measures the model components should give as output after simulation. These

**Table 3:** This table shows the design space matrix (DSM) for the case study, and it highlights one possible design.

| | weigh1.in | weigh2.in | weigh3.in | weigh4.in | weigh5.in | assign1.in | assign2.in | assign3.in | assign4.in | assign5.in | trim1.in | trim2.in | trim3.in | trim4.in | trim5.in | distribute1.in | distribute2.in | distribute3.in | distribute4.in | distribute5.in | distribute6.in | distribute7.in | distribute8.in | batchingDestination1.in | batchingDestination2.in | burgerDestination.in | schnitzelDestination.in | filletStripsDestination.in | mincedMeatDestination.in | Min. connections | Max. connections |
|---|---|---|---|---|---|---|---|---|---|---|---|---|---|---|---|---|---|---|---|---|---|---|---|---|---|---|---|---|---|---|---|
| origin1.out | ✓̄ | | | | | | | | | | | | | | | | | | | | | | | | | | | | | 1 | 1 |
| origin2.out | | ✓̄ | | | | | | | | | | | | | | | | | | | | | | | | | | | | 1 | 1 |
| origin3.out | | | ✓̄ | | | | | | | | | | | | | | | | | | | | | | | | | | | 1 | 1 |
| origin4.out | | | | ✓̄ | | | | | | | | | | | | | | | | | | | | | | | | | | 1 | 1 |
| origin5.out | | | | | ✓̄ | | | | | | | | | | | | | | | | | | | | | | | | | 1 | 1 |
| weigh1.out | | | | | | ✓̄ | | | | | | | | | | | | | | | | | | | | | | | | 1 | 1 |
| weigh2.out | | | | | | | ✓̄ | | | | | | | | | | | | | | | | | | | | | | | 1 | 1 |
| weigh3.out | | | | | | | | ✓̄ | | | | | | | | | | | | | | | | | | | | | | 1 | 1 |
| weigh4.out | | | | | | | | | ✓̄ | | | | | | | | | | | | | | | | | | | | | 1 | 1 |
| weigh5.out | | | | | | | | | | ✓̄ | | | | | | | | | | | | | | | | | | | | 1 | 1 |
| assign1.out | | | | | | | | | | | ✓ | | | | | ✓̄ | | | | | | | | | | | | | | 1 | 1 |
| assign2.out | | | | | | | | | | | | ✓ | | | | | ✓̄ | | | | | | | | | | | | | 1 | 1 |
| assign3.out | | | | | | | | | | | | | ✓̄ | | | | | ✓ | | | | | | | | | | | | 1 | 1 |
| assign4.out | | | | | | | | | | | | | | ✓̄ | | | | | ✓ | | | | | | | | | | | 1 | 1 |
| assign5.out | | | | | | | | | | | | | | | ✓ | | | | | ✓̄ | | | | | | | | | | 1 | 1 |
| trim1.out1 | | | | | | | | | | | | | | | | ✓ | | | | | | | | | | | | | | 0 | 1 |
| trim1.out2 | | | | | | | | | | | | | | | | | | | | | | | | | | | | | ✓ | 0 | 1 |
| trim2.out1 | | | | | | | | | | | | | | | | | ✓ | | | | | | | | | | | | | 0 | 1 |
| trim2.out2 | | | | | | | | | | | | | | | | | | | | | | | | | | | | | ✓ | 0 | 1 |
| trim3.out1 | | | | | | | | | | | | | | | | | | ✓̄ | | | | | | | | | | | | 0 | 1 |
| trim3.out2 | | | | | | | | | | | | | | | | | | | | | | | | | | | | | ✓̄ | 0 | 1 |
| trim4.out1 | | | | | | | | | | | | | | | | | | | ✓̄ | | | | | | | | | | | 0 | 1 |
| trim4.out2 | | | | | | | | | | | | | | | | | | | | | | | | | | | | | ✓̄ | 0 | 1 |
| trim5.out1 | | | | | | | | | | | | | | | | | | | | ✓ | | | | | | | | | | 0 | 1 |
| trim5.out2 | | | | | | | | | | | | | | | | | | | | | | | | | | | | | ✓ | 0 | 1 |
| distribute1.out1 | | | | | | | | | | | | | | | | | | | | | ✓ | ✓ | ✓̄ | ✓ | ✓ | | | | | 1 | 1 |
| distribute1.out2 | | | | | | | | | | | | | | | | | | | | | | | | | | | ✓̄ | | | 1 | 1 |
| distribute2.out1 | | | | | | | | | | | | | | | | | | | | | ✓ | ✓ | ✓ | ✓̄ | ✓ | | | | | 1 | 1 |
| distribute2.out2 | | | | | | | | | | | | | | | | | | | | | | | | | | | ✓̄ | | | 1 | 1 |
| distribute3.out1 | | | | | | | | | | | | | | | | | | | | | ✓ | ✓ | ✓ | ✓ | ✓̄ | | | | | 1 | 1 |
| distribute3.out2 | | | | | | | | | | | | | | | | | | | | | | | | | | | ✓̄ | | | 1 | 1 |
| distribute4.out1 | | | | | | | | | | | | | | | | | | | | | ✓̄ | ✓ | ✓ | ✓ | ✓ | | | | | 1 | 1 |
| distribute4.out2 | | | | | | | | | | | | | | | | | | | | | | | | | | | ✓̄ | | | 1 | 1 |
| distribute5.out1 | | | | | | | | | | | | | | | | | | | | | ✓ | ✓̄ | ✓ | ✓ | ✓ | | | | | 1 | 1 |
| distribute5.out2 | | | | | | | | | | | | | | | | | | | | | | | | | | | ✓̄ | | | 1 | 1 |
| distribute6.out1 | | | | | | | | | | | | | | | | | | | | | | | | | | ✓̄ | ✓ | | | 1 | 1 |
| distribute6.out2 | | | | | | | | | | | | | | | | | | | | | | | | | | | | ✓̄ | | 1 | 1 |
| distribute7.out1 | | | | | | | | | | | | | | | | | | | | | | | | | | ✓ | ✓̄ | | | 1 | 1 |
| distribute7.out2 | | | | | | | | | | | | | | | | | | | | | | | | | | | | ✓̄ | | 1 | 1 |
| distribute8.out1 | | | | | | | | | | | | | | | | | | | | | | | | | | ✓̄ | | | | 1 | 1 |
| distribute8.out2 | | | | | | | | | | | | | | | | | | | | | | | | | | | | ✓̄ | | 1 | 1 |
| Min. connections | 1 | 1 | 1 | 1 | 1 | 1 | 1 | 1 | 1 | 1 | 0 | 0 | 0 | 0 | 0 | 1 | 1 | 1 | 1 | 1 | 1 | 1 | 1 | 3 | 3 | 1 | 1 | 5 | 0 | | |
| Max. connections | 1 | 1 | 1 | 1 | 1 | 1 | 1 | 1 | 1 | 1 | 1 | 1 | 1 | 1 | 1 | 1 | 1 | 1 | 1 | 1 | 1 | 1 | 1 | 3 | 3 | 1 | 1 | 5 | 5 | | |

The rows represent all output ports of the DSM, the columns represent input ports. The check marks indicate which connection between input and output ports are allowed. The two columns to the right of the matrix and the two rows below the matrix indicate the minimum and maximum number of connections that each port is allowed to have. The green encircled check marks indicate the connections in the current system design shown in Figure 3.

performance measures are later used in the 'evaluation' step to score the system design on the performance indicators selected by the user of the design space exploration method.

### 3.2.1 Case study

For the case study, a model library was built in Anylogic. The model library consists of all model components for the plant modules such as *weighing* and *distribution* described in Section 2.3, and one model component for the production controller. These model components are built using Anylogic's process-modeling library. Each model component has a number of input and/or output ports. Between the input and output ports of a component, are a number of process-modeling blocks, which describe the production processes that are executed on the product, and their duration.

To model a complete system, specific instances of the model components can be placed on a canvas. These instances can be connected by their ports. Anylogic allows the model components to be connected to each other through the call of a function, allowing the same canvas to be used for all of the different designs. Figure 6 shows a simulation model constructed based on the design of the current system. This design is also highlighted in Table 3 and shown in Figure 3.

To correctly set up the model component of the production controller, the possible routings through the system have to be calculated for this specific system design. These routings can be deduced from the connections which were selected in the DSM, and from knowledge of how fillets flow through the modules. For example, for the model in Figure 6, it can be deduced that fillets in lane 3 can be routed to either the *batchingDestination2* or the *filletStripsDestination*, and that these fillets can be trimmed if necessary.

Besides the *mincedMeatDestination*, each destination corresponds to a recipe as defined in Table 2. Each destination model component records which percentage of the target throughput is fulfilled for the recipe corresponding to that destination. These are the performance measures later used in the 'evaluation' step.

## 3.3 Simulation

The third step in the methodology involves simulating the constructed model. In order to accurately predict the performance of a production system, it is necessary to simulate its behavior under a range of conditions, as highlighted by Penazzi et al. [7]. The operating conditions for which the production system is simulated are described in a production *scenario*. The selection of scenarios to be simulated is determined by the designer of the system. In rigid production systems, where operations are consistent, a single scenario may be sufficient to predict system performance. Conversely, in flexible manufacturing systems, a variety of production scenarios may be necessary to obtain a comprehensive understanding of system performance. Additionally, specific scenarios may be included to account for the introduction of new products or machine breakdowns.

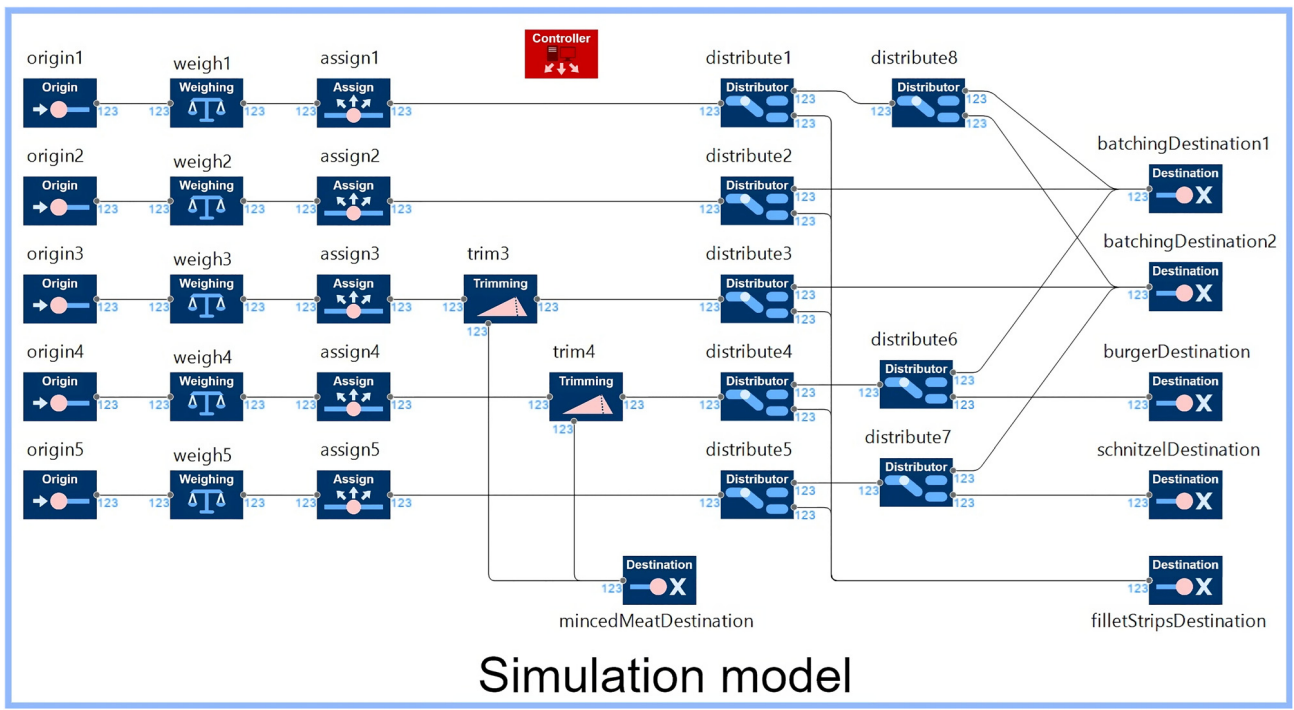

**Figure 6:** The simulation model of a system design as built after the 'model construction' step. This figure shows the simulation model of the current system design shown in Figure 3, which is also highlighted in the DSM in Table 3.

Besides selecting the simulation scenarios, simulation parameters must be chosen. Law [21] notes that when creating a simulation experiment it is necessary to tune simulation parameters such as simulation duration, the length of the warmup period, and the number of (independent) simulation runs. Law describes some methods on how these parameters can be tuned to find the right balance between the accuracy of the results and computation time. During each 'Simulation' step, the constructed model is simulated for the chosen simulation scenarios and parameters. The outcome of this step is the performance of the design under the specified scenarios. The performance measures which are obtained in this step were previously chosen during the construction of the model library.

### 3.3.1 Case study: selecting the simulation scenarios

As mentioned in Section 2.1, it is of vital importance that the system performs well for the many different incoming flocks. This requires that the effect of the mean fillet weight of a flock on the performance of the system is analyzed. Sensitivity analysis can be used to analyze the influence of a parameter on the system's performance [21]. A sensitivity analysis was carried out on the current system layout, to analyze how much influence the weight distribution of the incoming flow of fillets has on how much of the target throughput can be reached. The findings of this analysis were that the fillet weight distribution does indeed have a substantial effect on how much of the target throughput can be reached (a difference of up to 20 % was identified). To account for the variation in fillet weight distribution, five summer scenarios and five winter scenarios have been selected to be simulated in the case study, with mean fillet weights ranging from 200 to 300 g. Table 4 shows the selected scenarios. Increasing the number of scenarios further would result in a more accurate estimation of the variation in performance, with the downside that it would take a longer computation time to simulate all scenarios for every design.

**Table 4:** The 10 selected simulation scenarios. The summer and winter recipes are found in Table 2.

| Scenario | Recipes | Mean fillet weight (g) | Scenario | Recipes | Mean fillet weight (g) |
|---|---|---|---|---|---|
| 1 | Summer | 200 | 6 | Winter | 200 |
| 2 | Summer | 225 | 7 | Winter | 225 |
| 3 | Summer | 250 | 8 | Winter | 250 |
| 4 | Summer | 275 | 9 | Winter | 275 |
| 5 | Summer | 300 | 10 | Winter | 300 |

### 3.3.2 Case study: choosing the simulation parameters

Next, the simulation duration, the length of the warmup period, and the number of independent simulation runs were chosen. These parameters are tuned to find the right balance between accuracy of results and computation time. These three parameters were analyzed based on the simulation model of the current system (shown in Figure 3) and the most common production scenarios for both summer and winter (scenarios 3 and 8 from Table 4). The performance measure in these two scenarios is the average percentage of the target throughput achieved for the four main recipes shown in Table 2 (not including the fillet strips recipe). The goal was to find the right settings that give an accurate estimation of this performance measure for a low computation time.

First, the warmup period was chosen to be 200 s. The warmup period is needed because enough fillets need to be weighed before the production controller can make a production strategy. Increasing the warmup period beyond 200 s has shown to have no demonstrable effect on the accuracy of the simulation.

Next, an analysis was done on how the simulation duration affects the accuracy and precision in predicting the performance measures, and how it affects the computation time of the simulation. The simulation duration was varied between 400 and 25,600 s, and each setting was simulated 30 times. Figure 7 shows how the simulation duration affects the accuracy and precision in predicting the performance measures for scenarios 3 and 8. Calculating the precision and accuracy requires a reference point. Preferably, this would be obtained through experiments with the real-world system. However, this was not possible in this case study, so instead, the mean of 30 simulations with a duration of 25,600 s was taken as the reference.

In both scenarios, simulating 30 times with a simulation duration of 3200 s resulted in a mean within 0.01 % of the reference. This shows that the results are *accurate*; given enough simulations of 3200 s the mean will converge to that of the reference. All of the 30 simulations with a simulation duration of 3200 s were within 1 % of the mean. This shows that the results are also *precise*; the results of individual simulation runs are close to each other. Combined, these indicators show that one simulation run of 3200 s should give accurate results which are likely to be within 1 % error. However, there is no guarantee as these results are based on the current system, and for only two scenarios. For any other design or scenario, the same simulation parameters could be inadequate. A solution to this problem is to repeat this experiment with the recommended designs to validate the accuracy of the simulation results.

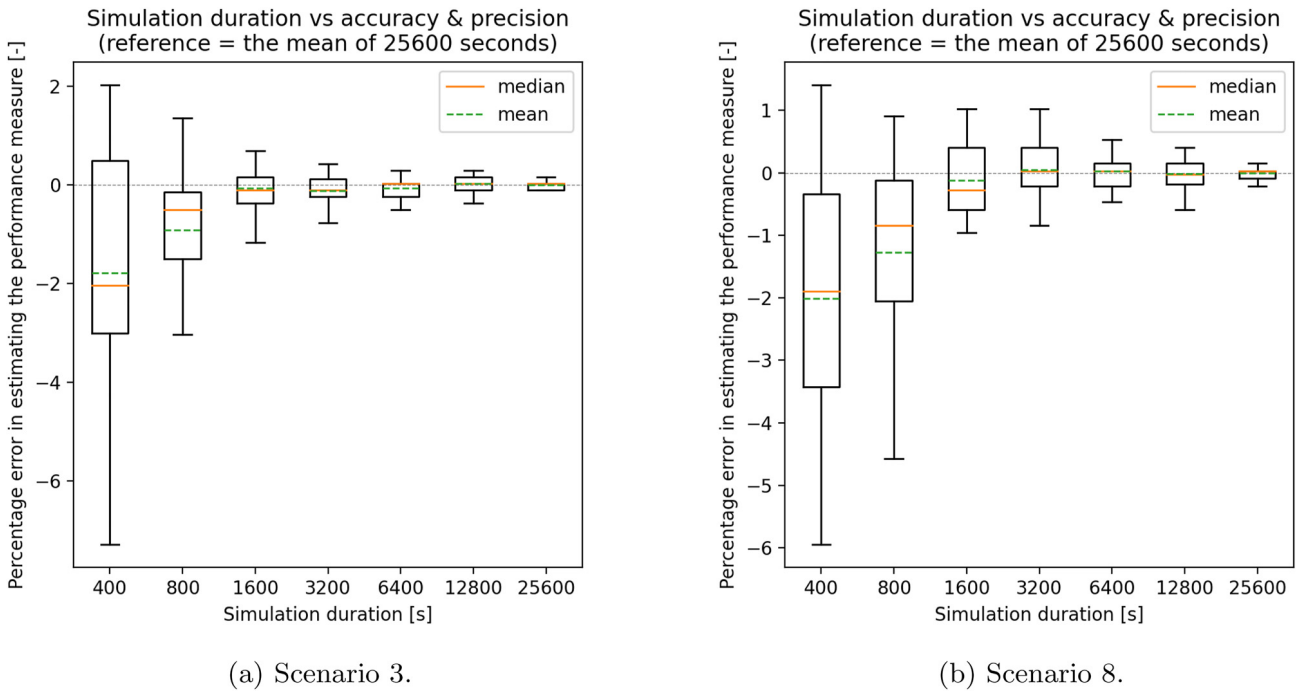

(a) Scenario 3.

(b) Scenario 8.

**Figure 7:** Figures (a) and (b) show a 'boxplot'. These figures show how the simulation duration affects the accuracy and precision in predicting the performance measure in scenarios 1 and 2. The vertical axes show the percentage error in estimating the performance measure (percentage of target throughput realized). The mean performance measure of all simulations with a duration of 25,600 s is taken as the reference. A simulation duration produces accurate results if its mean is close to the reference. A simulation duration produces precise results if its estimations are close to each other (a small 'box' and 'whiskers').

Finally, the expected total computation time for simulating one independent simulation run of 3200 s for all scenarios (10) and each design (11,520), was calculated to be less than one day. This was deemed to be a good balance between accuracy and computation time for the purpose of this case study.

### 3.3.3 Case study: total computation time

For each design, the constructed model was simulated once for each of the chosen scenarios. All simulations were done for 3200 s of production, during which a total of around 18,000 fillets are processed. In total, simulation of all 11,520 designs, 10 scenarios each, resulted in around 12 years of production being simulated. This took roughly half a day of computation, using a PC with an Intel(R) Core(TM) i5-8365U CPU with 16GB of RAM.

## 3.4 Evaluation

In the final stage of the design space exploration method, the performance of the designs is evaluated. This evaluation is conducted using performance indicators selected by the system designer. Which performance indicators are relevant largely depends on the chosen production scenarios. Different types of performance indicators might be required for different scenarios. Only performance measures included in the construction of the model library can be used in the calculation of these performance indicators.

Once a design has been evaluated, the outcome is recorded, and the process proceeds to the next iteration. This continues until all designs have been evaluated or until a predefined stopping criterion has been met (e.g., the designer only seeks the first design that satisfies the specified requirements). Ultimately, the method yields one or more recommended designs based on their scores on the selected performance indicators. However, there are many methods for selecting which design(s) to recommend. Two methods for single- and multi-objective optimization are described below, and examples for both are given in the case study.

The first proposed method is to use single-objective optimization. A fillet processing system has many relevant performance measures. When there are many different performance measures, it can become difficult to identify which designs are best. In such cases, an effective approach can be to collect all performance measures into a single objective function. The goal is then to find the design(s) that score best according to this function.

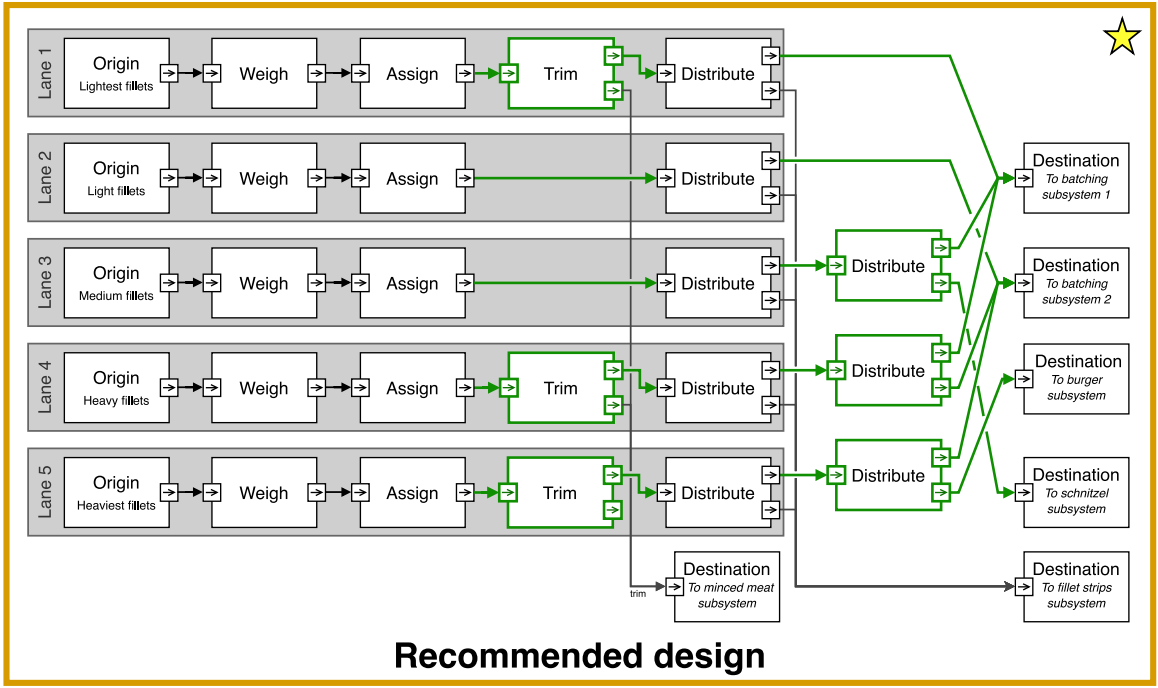

**Figure 8:** The design space exploration method recommends the design shown as it has the highest 10-year ROI. In the design, the selected connections and modules are depicted in green.

Another option for selecting the best design is to use multi-objective optimization to identify all designs that are Pareto optimal. A design is Pareto optimal if no other design scores better in at least one performance indicator, without having to sacrifice in another performance indicator [22]. Multiple designs can be Pareto optimal, each offering a unique trade-off between the selected performance measures. The advantage of this approach is that the system designer can select which of the Pareto optimal designs offers the best trade-off.

Besides optimization, this design space exploration method can also be used for extensive analysis of the design space. The case study shows two examples of how the collected performance data can be used to answer specific design questions.

### 3.4.1 Case study: single-objective optimization

The single-objective optimization approach is to select the design that scored best on a chosen objective function. For example, suppose that the goal is to find the design with the highest 10-year return on investment (ROI), with the ROI being calculated as follows:

$$10-\text{year ROI} = \frac{(w+s) \cdot P \cdot Y}{B + t \cdot M} \cdot 100\,\%$$

With the following variables:
(A) The average percentage of target throughput reached in summer scenarios over all recipes: $s$.
(B) The average percentage of target throughput reached in winter scenarios over all recipes: $w$.
(C) The number of trimming modules: $t$.

And constants:
(I) Each percentage of throughput in summer and in winter results in the yearly profit: $P = \$10000$.
(II) The number of years over which ROI is calculated: $Y = 10$ years.
(III) The base cost of the system: $B = \$10$ million.
(IV) The cost of each trimming module: $M = \$50000$.

When the design space exploration method is used to optimize towards the 10-year ROI, then the method recommends the design shown in Figure 8. This design has the highest 10-year ROI at 167 %.

### 3.4.2 Case study: multi-objective optimization

The second approach is to use multi-objective optimization, in which the goal is to identify all Pareto optimal designs. For this approach, the three variables (A), (B), and (C), which were also used in the calculation of the single-objective cost function, are chosen as objectives.

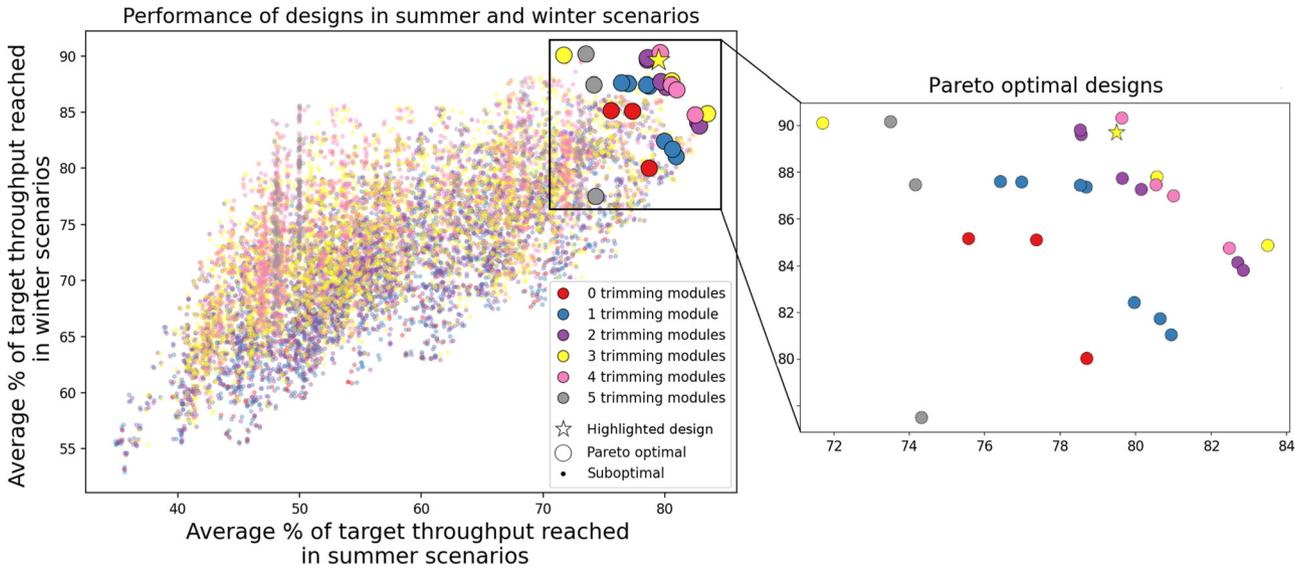

**Figure 9:** The performance for all of the 11,520 different designs. Each dot represents one design; designs that are Pareto optimal with respect to (A) and (B) are indicated with a bigger dot. The right figure zooms in on these Pareto optimal designs. The Pareto optimal design which has the highest 10-year ROI is highlighted with a star, and is shown in Figure 8.

Figure 9 shows the performance of each of the 11,520 designs. In the figure, each dot represents one design. The performance in summer scenarios (A) and winter scenarios (B) is displayed on the horizontal and vertical axes, respectively. The color of the dot indicates the number of trimming modules (C). When judging the designs on these three performance indicators, only the Pareto optimal designs are relevant (given that your simulation is accurate enough). For each number of trimming modules (C), the Pareto optimal designs that score best in summer (A) and winter (B) are highlighted with a bigger dot. These designs are in the top right corner of the figure. The right figure zooms in on these 27 Pareto optimal designs. Also highlighted in the figure with a star is the (Pareto optimal) design depicted in Figure 8, which had the highest 10-year ROI according to the single-objective cost function.

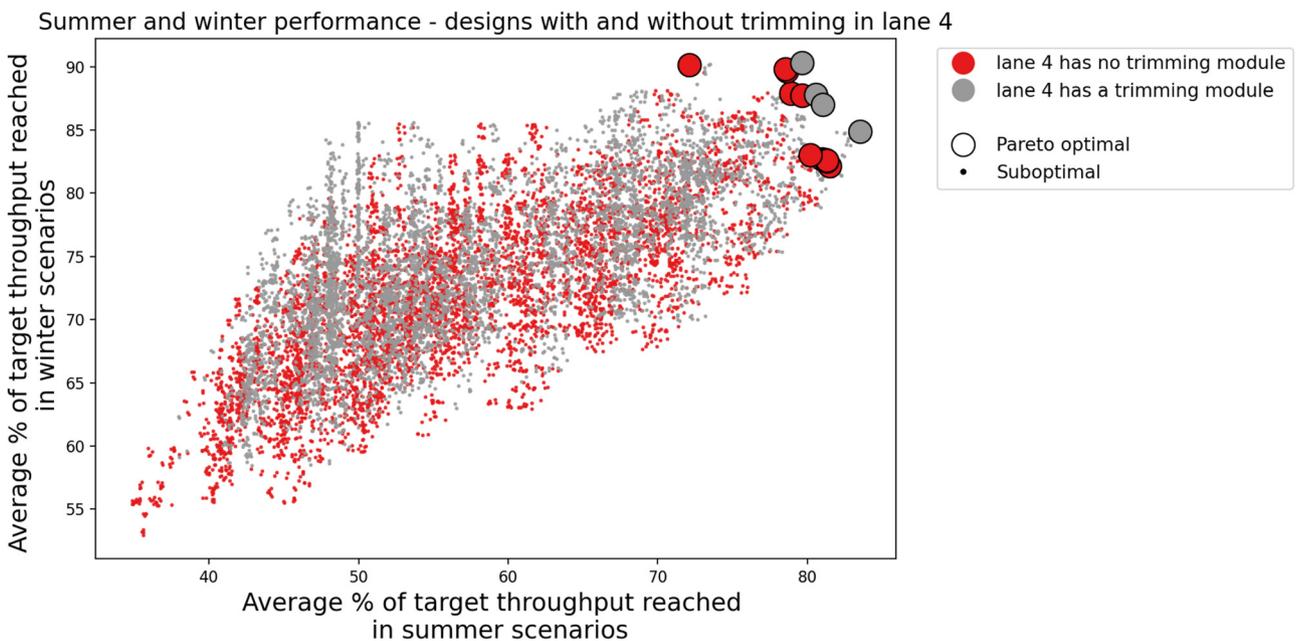

**Figure 10:** The performance in summer and winter of systems with or without a trimming module in lane 4.

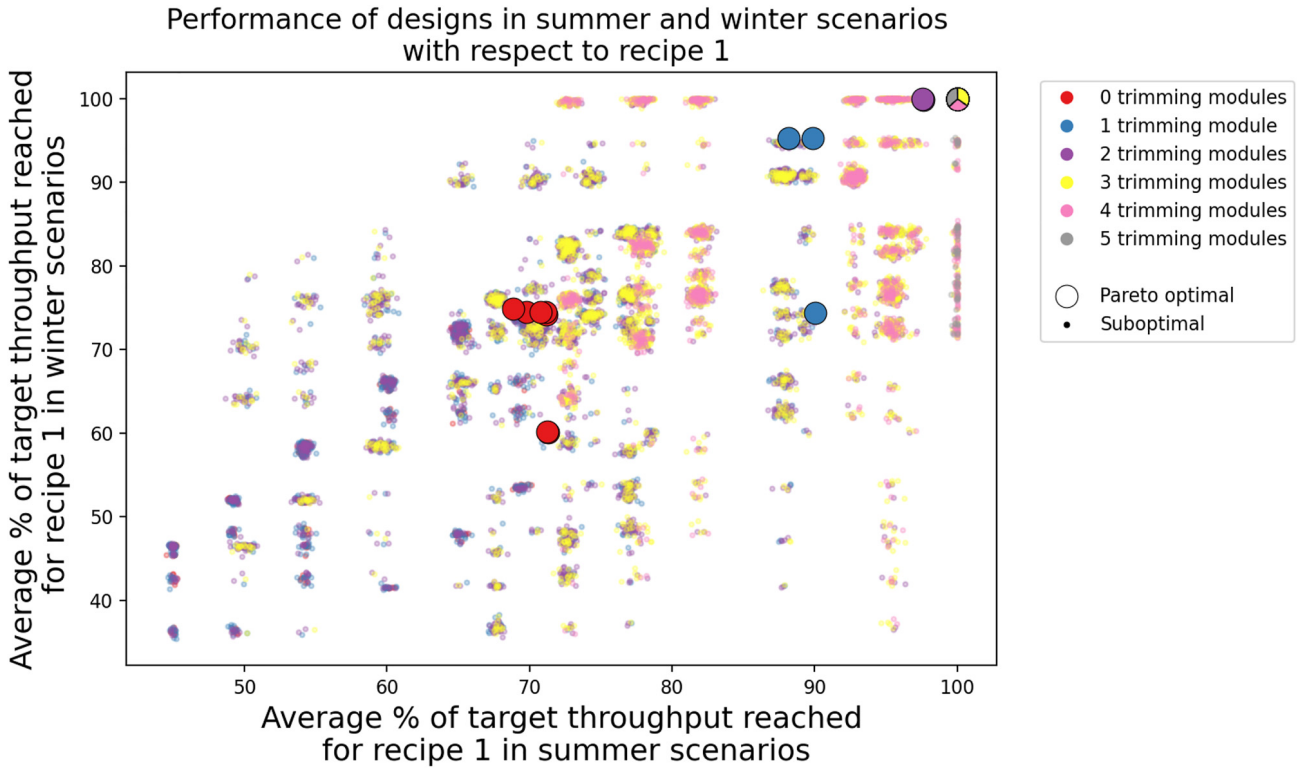

**Figure 11:** The performance of each design with respect to recipe 1. Pareto optimal designs with 3, 4, and 5 trimming modules in the top right corner overlap, which is why the marker shows three colors.

A system designer can use Figure 9 to select which design is most suited to the needs of the project. The designs with the highest performance are in the top right of the figure; these are the designs with three or four trimming modules. The poor performance for systems with five trimming modules might be unexpected as more trimming modules should give more flexibility, and thus better results (as it is always possible not to use the trimming module). However, Algorithm 1 for calculating the production strategy is rather naive, as it prioritizes the throughput of recipes with a higher priority over getting an optimal overall throughput. One takeaway of this analysis might be that a better algorithm needs to be developed.

### 3.4.3 Case study: further analysis of the design space

This section shows how the collected performance data can be used to further analyze the design space with regard to specific design questions. An example of such a design question is whether lane 4 should or should not have a trimming module. Figure 10 shows which designs perform better in performance measures for summer (A) and winter (B): designs with, or without a trimming module in lane 4. As can be established from the figure, designs with a trimming module in lane 4 outscore designs without one. Note that the number of trimming modules is not considered, resulting in a different set of Pareto optimal designs.

Similarly, it is possible to answer design questions with respect to specific performance measures. Previously, the performance was average over all of the four recipes. Suppose that the system designer wants to analyze which designs are best for fulfilling the target throughput of only recipe 1.

Figure 11 shows the performance of each design with respect to recipe 1. As can be seen in the figure, a design with at least 3 trimming modules is required to obtain a 100 % throughput for recipe 1 in all winter and summer scenarios. The 'clustering' in Figure 11 is due to design choices being discrete. There is a limited number of ways to connect the incoming lanes to the destination of recipe 1. The difference in performance between designs in a cluster is due to random variation in the weights of fillets.

## 4 Discussion

In this section, each of the four steps in the design space exploration method are discussed.

## 4.1 Discussion on design space exploration

The Design Space Matrix, while useful in describing the design space of a production system, may not always be sufficient for capturing the complexity of certain design specifications. One such example is the requirement: if component A is connected to component B, component C must be connected to component D. Furthermore, the design space may contain nonsensical or impermissible designs, such as the creation of a loop in which a destination can never be reached. Also, the current implementation only allows changing connections or swapping modules in or out. It does not allow for the parameters of these modules (e.g., their processing time) to be changed. One potential solution to these issues is to use a specification language that can be used to describe the design space, such as proposed in [24].

Another problem is that functionally equivalent designs may exist within the design space, resulting in a reduction in the number of distinct designs. It is important to consider pruning the design space of functionally equivalent designs [23].

The current approach for exploring the design space, which involves iterating through various configurations, is limited in its efficiency and effectiveness. Some alternative methods for more intelligent exploration of the design space, such as simulated annealing or genetic algorithms, are proposed as potential solutions in Pimentel [23].

## 4.2 Discussion on model construction

The advantage of using a library for model construction is its reusability. Model components are reused in all the different designs, and can even be repurposed in future case studies. However, the main challenge in model construction is that building a model library requires very specific expertise.

One of the challenges in constructing the model library lies in creating a model for the production control layer. In the case of a fillet processing system, the production controller uses a generic design that can be applied to any plant layout. This is possible because the production controller always interacts with the other components in the same way. Adapting the production controller to a new design only requires all possible product routings to be calculated.

However, more research is needed to determine how the proposed model construction method generalizes to other types of production systems. The method works best for systems in which the production control layer is flexible enough to deal with different designs. The proposed method might not be feasible for systems in which the production control layer needs to be tailor-made for each design.

## 4.3 Discussion on simulation

The bottleneck of this method is in the computation time for simulating all designs. In the case study, simulating 12 years of production, across 11,520 different designs and 10 scenarios, required roughly half a day of computation. However, when dealing with bigger design spaces, or with more complex models, iteratively simulating all designs might be infeasible. This makes it crucial to implement intelligent methods for exploring the design space, in which not all designs need to be simulated.

Still, even if intelligent methods for exploring the design space are implemented, scalability can remain a challenge for larger and more complex systems. The simulation of even a small number of very complex designs can be computationally expensive. Xu et al. [22] suggests that utilizing multi-fidelity simulation, which utilizes models of varying fidelity levels, may help reduce computational costs by utilizing lower-fidelity models when many design alternatives are being considered.

## 4.4 Discussion on evaluation

As shown in Section 3.4, the evaluation of designs can be accomplished through various methods. One method is multi-objective optimization, which can be used to identify Pareto optimal designs. This enables the designer to select the preferred design from the available options. However, when more than three performance indicators are present, visualizing the system's performance becomes challenging, which may impede the designer's ability to interpret the results and select the best option. In such cases, it is recommended to employ an alternative approach that allows the method to determine the recommended designs, such as single-objective optimization.

Defining the right objective function is not always straightforward, particularly when comparing the values of different types of performance indicators. It becomes even more difficult when scenarios that are completely different must be compared. For instance, it is difficult to compare a scenario where the system operates under normal conditions with a scenario in which a machine failure occurs. One solution to this challenge is to establish minimum requirements that a design must meet, such as a minimum of 80 % target throughput for normal conditions and 60 % for the case of a breakdown. Another solution could be to assign weights to the different performance indicators, such as assigning a weight of 0.9 to performance under normal production and a weight of 0.1 to performance during machine failure.

# 5 Concluding remarks

This paper presents a method for design space exploration using discrete-event simulation in which most of the steps are automated. This greatly reduces the time and effort required to iterate through different designs. This method takes as input the specified design space of the production system, a model library with model components for each of the system's modules, and the chosen production scenarios and performance indicators. The method has four steps: design space exploration, model construction, simulation, and evaluation. Most of the effort when using this method is in constructing the model library. However, one of the biggest advantages is that this model library can be reused for future case studies. Multiple approaches are presented for evaluating the different designs in the design space. It is shown how the system designer can use this method for both single- and multi-objective optimization, and how it can be used to answer specific design questions.

The method is validated through a case study of industrial-level complexity. In this case study 11,520 different designs are compared over 10 different simulation scenarios. The presented case study shows that this method is effective for design space exploration of poultry processing systems. Our hypothesis is that these results extend to other types of food processing systems, and production systems in general.

One of the main challenges of this method is in dealing with case studies of increased complexity; the bottleneck of the method is in simulating the many different system designs for multiple scenarios. An improvement to the proposed design space exploration method would be to iterate through the designs more intelligently, which could be achieved by using optimization methods, and/or by using feedback from the 'evaluation' step to identify which direction design space exploration should continue. If needed, the computational cost required for simulation could be reduced by using multi-fidelity simulation.

Another challenge lies in the specification of the design space. In this work, the design space is specified through a design space matrix, which denotes which connections are, and are not allowed in a design. This method is sufficient for the case study presented in this paper, but it is quite limited in terms of expressiveness. In future work, the authors would like to integrate the more expressive design space specification language proposed in Paape et al. [24] into the design space exploration method of this work.

**Research ethics:** Not applicable.
**Author contributions:** The authors have accepted responsibility for the entire content of this manuscript and approved its submission.
**Competing interests:** The authors state no conflict of interest.
**Research funding:** None declared.
**Data availability:** Not applicable.